# Antiphase boundary-driven large exchange bias and negative magnetoresistance in $NiCo_2O_4$

Biswanath Pramanik, Pushpesh Pathak, and Binoy K. Hazra*

Department of Physics, Indian Institute of Technology Guwahati, Assam 781039, India

*bkhazra@iitg.ac.in

## ABSTRACT

Nickel-based spinel oxide has recently attracted significant attention due to its remarkable magneto-transport properties, promising various spintronic applications. In this study, we observe an exchange bias effect, where the magnetic hysteresis loop shifts both horizontally and vertically when single-phase nanocrystalline $NiCo_2O_4$ is cooled in the presence of an external magnetic field. The magnitude of the exchange bias field is approximately 375 Oe at 5 K, and this effect disappears around 300 K. Furthermore, $NiCo_2O_4$ shows semiconducting behaviour in longitudinal resistivity and demonstrates a substantial negative magnetoresistance of ∼31.5% at 10 K. The magnetoresistance exhibits a butterfly-shaped behaviour at low magnetic fields and becomes almost linear at high magnetic fields. A detailed analysis of the negative magnetoresistance and exchange bias reveals that both phenomena originate from the formation of antiphase boundaries in the as-synthesized $NiCo_2O_4$ sample, as confirmed by the high-resolution transmission electron micrograph.

## 1. Introduction

Transition metal oxides have garnered extensive attention owing to their intriguing magnetic behaviour, diverse magneto-transport properties, and wide-ranging potential for energy storage and spintronic applications [1-3]. Among them, nickel cobaltite ($NiCo_2O_4$) stands out as a mixed-valence spinel oxide exhibiting remarkable electrical and magnetic versatility. In nanoparticle form, $NiCo_2O_4$ (NCO) has been widely investigated for its exceptional performance in supercapacitors [4], lithium-ion batteries [5], electrocatalysts [6], fuel cells [7], and sensors [8,9]. Despite these advances, the magnetic and magneto-transport properties of bulk NCO remain relatively underexplored. Recent studies on NCO thin films have revealed strong out-of-plane magnetic anisotropy and a pronounced anomalous Hall effect (AHE) that persists well above room temperature, underscoring its potential in spintronic device architectures [10]. With a Néel temperature of approximately 400 K and a ferrimagnetic ground state exhibiting low saturation magnetization, NCO emerges as a highly promising material for next-generation spintronic and multifunctional oxide-based technologies [10, 11].

Among the various magnetic phenomena relevant to spintronic functionality, exchange bias (EB) is a fascinating phenomenon typically observed at the interfaces between ferromagnetic (FM) and antiferromagnetic (AFM) materials, as well as in ferrimagnets [12,13]. The observation of a finite EB in a single-phase material is particularly intriguing, as it provides a simpler and more efficient alternative to conventional multilayer architectures in spintronic devices [14,15]. In such systems, the EB field can serve as an intrinsic bias field, enabling spin-orbit torque (SOT)-driven magnetization switching or spin-transfer torque (STT)-driven magnetic tunnel junction (MTJ) operations [15]. Finite EB has been reported in bilayer systems involving NCO, such as NCO/NiO [16], where the interfacial unidirectional anisotropy between NCO and NiO is primarily responsible for the observed shift in hysteresis. More recently, a pronounced EB has also been observed in single-layer NCO thin films, attributed to the formation of an antiferromagnetic rock-salt phase ($Ni_xCo_{1-x}O$) at the interface between NCO and the $Al_2O_3$ substrate [17]. In contrast, only one study has reported EB in nanocrystalline bulk NCO, which was annealed for 12 hours, revealing a relatively weak EB field of about 42 Oe at 10 K. Notably, as-synthesized NCO, which was annealed for 2 hours, does not show EB. The same article reported significant negative magnetoresistance (MR) at low magnetic fields, arising from the fast rotation of magnetization in core nanocrystals of the as-synthesized sample . Further, the MR decreased with longer annealing time, which was correlated with intergrain tunnelling and antiphase boundaries (APBs) [18]. The magnetoresistance was found to be proportional to the square of the normalized magnetization at very low fields only for both the as-synthesized and annealed samples, but it deviates significantly at high fields across the entire temperature range.

Here, we have explored magnetic and magneto-transport investigations on as-synthesized single-phase NCO nanoparticles, which reveal captivating interfacial phenomena. A symmetric magnetic hysteresis has been observed when the NCO is cooled in the absence of an external magnetic field. In contrast, magnetic field-cooled magnetization measurements exhibit pronounced EB, characterized by distinct horizontal and vertical shifts in the magnetic hysteresis loops, indicating the presence of unidirectional anisotropy and uncompensated pinned spins within the NCO. In addition, a large negative MR is observed at 10 K, which diminishes with increasing temperature, and displays a characteristic butterfly-shaped field dependence at low magnetic fields. The MR further scales with $A_i \left( {}^{M}/_{M_s} \right)^{2i}$ $(i = 1, 2, 3)$ in a wide temperature range, indicating the presence of antiphase boundary which is supported by the high-resolution transmission electron micrographs. The antiphase boundary results in antiferromagnetic or frustrated exchange coupling between two grains, which is responsible for the observation of significant exchange bias in the NCO sample.

## 2. Experimental Details

NCO nanoparticles were synthesized using a cost-effective sol-gel combustion method. In this process, 1.22 g of $Ni(NO_3)_2 \cdot 6H_2O$ ( ≥ 99.9 % Thermo Fisher Scientific) and 2.44 g of $Co(NO_3)_2 \cdot 6H_2O$ ( ≥ 99.9 % Thermo Fisher Scientific ) were separately dissolved in 10 ml and 20 ml of deionized (DI) water to prepare 0.02 M and 0.04 M solutions, respectively. The two precursor solutions were then combined and magnetically stirred at 350 rpm for 30 minutes, maintaining a temperature of 80 °C to ensure complete dissolution and homogeneous mixing of the metal ions [19]. Subsequently, a 0.024 M aqueous solution of citric acid was added dropwise to the mixed nitrate solution. Citric acid acted as a chelating agent, promoting complexation between the metal cations and facilitating uniform gel formation. The reaction mixture was continuously heated on a heater-cum stirrer at 120 °C until a viscous stable gel was obtained. The resulting gel was then dried to form a pink solid, which was gently ground using an agate mortar and pestle to obtain a fine powder. To induce crystallization and achieve the formation of single-phase spinel-structured NCO, the dried precursor powder was calcined at 350 °C for 4 hours in a muffle furnace. The resulting black powder was subsequently used for further characterization and analysis.

The crystal structure of the synthesized sample was examined using a powder X-ray diffractometer (XRD, Rigaku Smartlab 9 kW) operated at 40 kV and 125 mA with Cu-Kα radiation (λ = 1.5406 Å) in Bragg-Brentano geometry. Further confirmation of the spinel structure was obtained from selected area electron diffraction (SAED) patterns and high-resolution micrographs recorded using a field emission transmission electron microscope (FETEM, JEOL-2100F). Elemental composition and microstructural features were analyzed by field emission scanning electron microscopy (FESEM, ZEISS Sigma 300) equipped with an energy-dispersive X-ray spectroscopy (EDS, ZEISS Sigma) unit, operated in SE2 detector mode at an accelerating voltage of 20 kV. X-ray photoelectron spectroscopy (XPS, PHI 5000 Versa Probe III, Physical Electronics) was employed to investigate the surface chemical composition and cationic states. Temperature-dependent magnetization and magneto-transport properties were measured using a Physical Property Measurement System (PPMS, Quantum Design Dynacool 9T) equipped with standard measurement accessories.

## 3. Results and discussion

### 3.1. Structural properties

Figure 1(a) shows the X-ray diffraction (XRD) pattern along with the Rietveld refinement. The Rietveld refinement of the diffraction pattern has been performed using the FullProf Suite software with the space group $Fd\overline{3}m$ (ICDD PDF 04-021-8094). The calculated diffraction peaks correspond well with the experimental data, indicating the reliability of the refinement process, which results in a low value of the goodness-of-fit parameter, *i.e.*, $\chi^2 = 1.48$, confirming the single-phase nature of spinel NCO nanoparticles without any impurity phases. The absence of secondary peaks indicates excellent structural order, consistent with earlier reports on the NCO nanoparticles prepared through similar synthesis routes [20]. The lattice parameter ($a$) is found to be 8.11 Å from the Rietveld refinement, which is in good agreement with the cubic spinel NCO structure [21, 22].

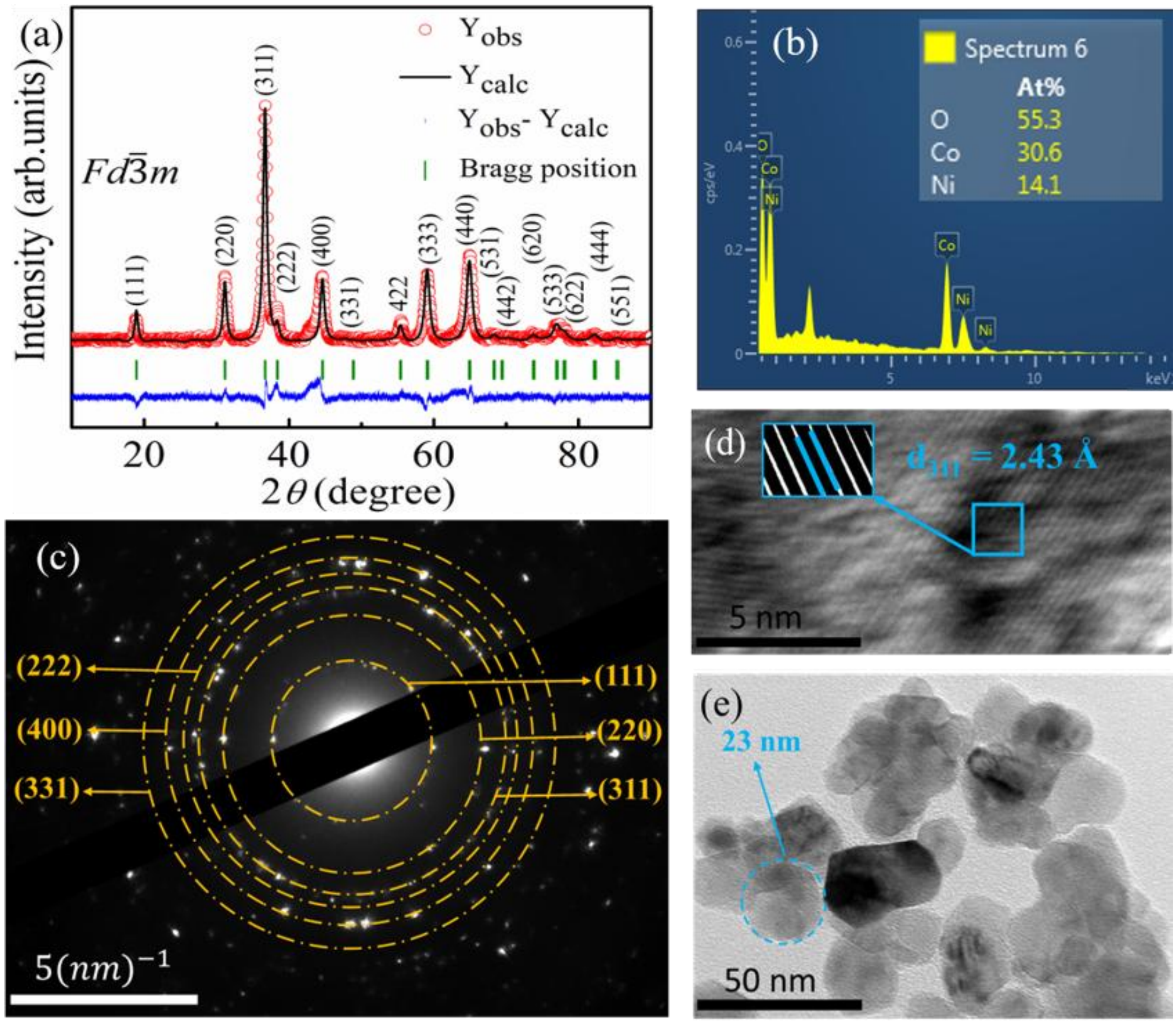

FIG. 1. (a) Rietveld refined XRD pattern of NCO using space group 227 [$Fd\bar{3}m$]. (b) Typical EDS spectrum displaying the atomic percentage of each constituent element in NCO. (c) SAED pattern of NCO, (d) HRTEM micrograph portraying (311) plane with inverse fast Fourier transform image as inset. (e) Bright field micrograph depicting the NCO in nanoparticle form.

The average crystallite size ($D_v$) was estimated using the Debye-Scherer equation [23], $D_v = {k\lambda}/{\beta cos\theta}$ where λ is the X-ray wavelength, $k$ is the shape factor (0.9), and $\beta$ represents the full width at half maximum (FWHM) of the most intense peak (311) after correction for instrumental broadening [23, 24]. The calculated $D_v$ of 13 ± 1 nm confirms the nanoscale dimensions of the synthesized NCO particles, which is further supported by the broadening of the primary diffraction peaks, a characteristic feature of nanocrystalline materials [23]. The energy-dispersive X-ray analysis reveals the atomic percentages of Ni:Co as ~ 1:2, which is in good agreement with the nominal stoichiometry of NCO. The corresponding EDX spectrum, illustrating the elemental distribution in the prepared nanoparticles, is presented in FIG. 1. (b). To gain a deeper understanding of the morphology, nanostructure, and crystallinity, FETEM analysis has been carried out. The corresponding selected area electron diffraction (SAED) pattern, displayed in FIG. 1. (c), reveals a series of well-defined bright diffraction rings composed of discrete spots, confirming the polycrystalline nature of the NCO nanoparticles. The calculated interplanar spacings ($d_{hkl}$) values, obtained from the measured ring diameters, correspond to the (111), (220), (311), (222), (400), and (331) crystallographic planes, which are characteristics of the cubic spinel structure of NCO [25]. The absence of any additional diffraction spots or rings further confirms the single-phase nature of the sample. FIG. 1. (d) presents the high-resolution transmission electron microscopy (HRTEM) image, clearly revealing the well-resolved lattice fringes corresponding to the (311) plane with an interplanar spacing ($d_{311}$) of 2.43 Å. The inset shows the inverse fast Fourier transform (ifFT) of the micrograph, providing enhanced visualization of the lattice fringes. The bright-field FETEM micrograph (FIG. 1. (e) reveals that the NCO nanoparticles consist of nearly spherical to slightly agglomerated nanocrystallites with uniform contrast, indicating good crystallinity. The average particle size is estimated to be around 23 ± 1 nm. The observed agglomeration of nanoparticles can be attributed to their high surface energy and magnetic interactions, which are common in nanoscale magnetic spinel oxides [25].

Figure 2(a) demonstrates the unit cell of the inverse spinel-structured NCO, comprising eight tetrahedral ($T_d$) and sixteen octahedral ($O_h$) interstitial sites available for cation occupancy. In this configuration, the $T_d$ sites are primarily occupied by $Co^{3+}$ ions, along with a fraction of $Co^{2+}$, reflecting their mixed-valence nature. Half of the $O_h$ sites are occupied by Ni ions, exhibiting mixed valence states ($Ni^{2+}/Ni^{3+}$), while the remaining $O_h$ sites are predominantly filled by $Co^{3+}$ ions [10,11]. To investigate the cation distribution of NCO nanoparticles, XPS analysis has been performed. The XPS spectra of Ni 2*p* and Co 2*p* are presented in FIG. 2. (b) and FIG. 2. (c), respectively. The Ni 2*p* spectrum exhibits two prominent peaks at 855.10 eV and 872.75 eV, while the Co 2*p* spectrum shows peaks at 780.02 eV and 795.32 eV, corresponding to the $2p_{3/2}$ and $2p_{1/2}$ with spin-orbit splitting of approximately ∼17.65 eV for Ni and ∼15.3 eV for Co, respectively [26, 27] . Deconvolution of the Ni 2*p* spectrum reveals $Ni^{3+}$ components at 853.44 eV and 866.93 eV, and $Ni^{2+}$ components at 855.42 eV and 871.29 eV, indicating a dominant $Ni^{3+}$ fraction ($Ni^{3+}/(Ni^{2+} + Ni^{3+}) \approx 0.80$). Similarly, the Co 2*p* spectrum displays $Co^{3+}$ peaks at 779.38 eV and 794.50 eV, and $Co^{2+}$ components at 780.74 eV and 796.07 eV, with $Co^{2+}$ being predominant ($Co^{2+}/(Co^{2+} + Co^{3+}) \approx 0.66$). Satellite peaks observed at 861.03 eV and 879.42 eV for Ni 2*p*, and at 785.15 eV and 802.90 eV for Co 2*p*, further confirm the characteristic features of transition-metal oxides [28, 29]. The O 1*s* spectrum (FIG. 2. (d)) can be deconvoluted into three distinct components located at 528.79 eV ($O_3$), 530.12 eV ($O_2$), and 532.94 eV ($O_1$), corresponding to lattice oxygen, surface oxygen with low coordination, and hydroxyl or adsorbed water species, respectively [30]. These XPS results confirm the presence of a mixed-valence inverse spinel structure, which can be expressed as:

$[Co^{2+}_{x}Co^{3+}_{1-x}]_{T_d}[Co^{3+}Ni^{2+}_{1-x}Ni^{3+}_{x}]_{O_h}O_4$ demonstrating the coexistence of $Co^{2+}/Co^{3+}$ and $Ni^{2+}/Ni^{3+}$ states, in excellent agreement with earlier reports [31].

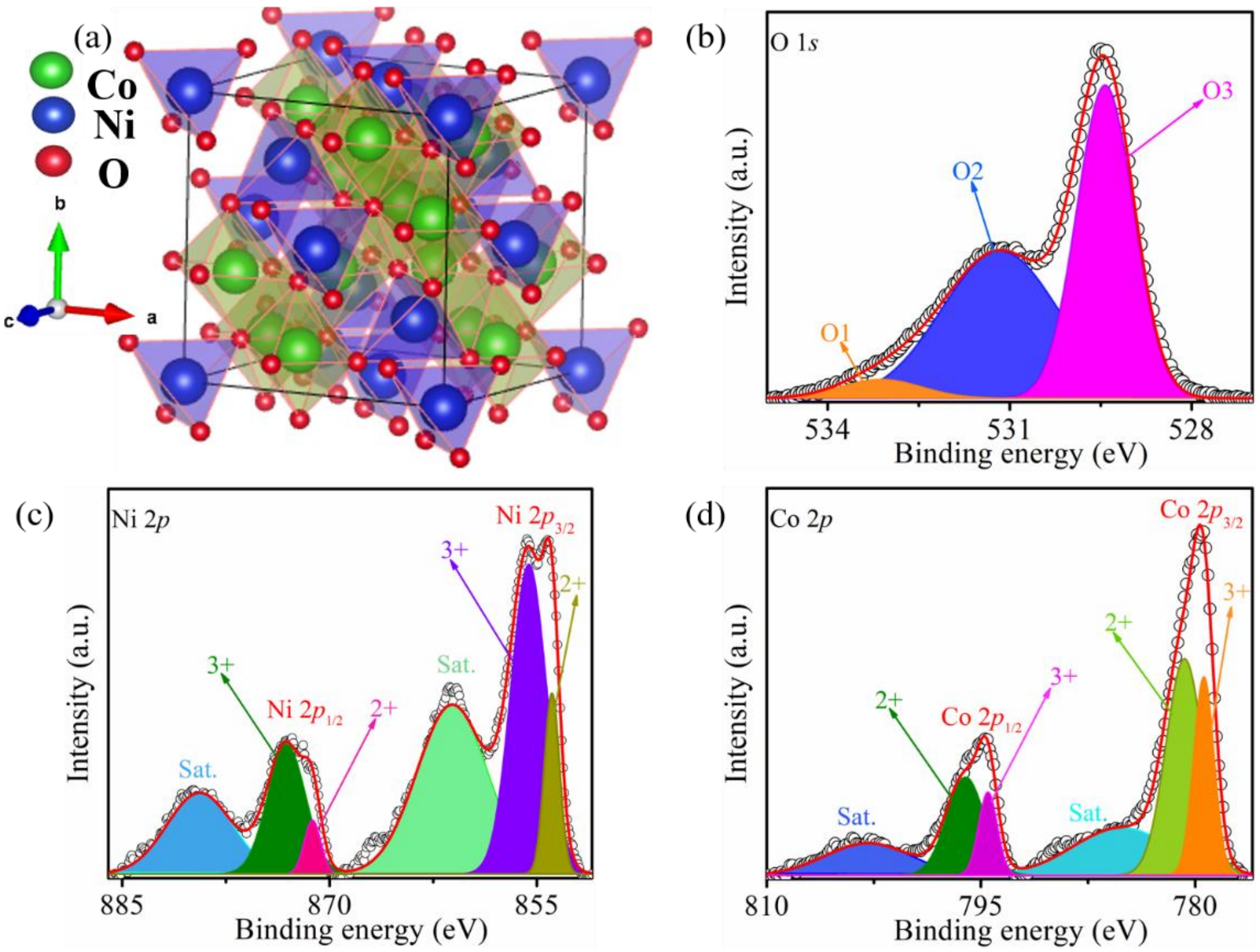

FIG. 2. (a) Inverse spinel crystal structure of NCO illustrating cation distribution within the unit cell. (b-d) XPS spectra of Ni 2*p*, Co 2*p*, and O 1*s* with Gaussian fittings, confirming the coexistence of $Ni^{2+}/Ni^{3+}$ and $Co^{2+}/Co^{3+}$ oxidation states, along with lattice, surface, and hydroxyl oxygen species, consistent with the mixed-valence inverse spinel structure of NCO.

### 3.2. Magnetic properties

To investigate the magnetic behaviour of NCO nanoparticles, temperature-dependent magnetization (*M*-*T*) measurements have been carried out under zero-field-cooled (ZFC) and field-cooled (FC) conditions with an applied magnetic field of 100 Oe, as shown in FIG. 3. (a). A distinct bifurcation between the ZFC and FC curves appears below 300 K, indicating the onset of magnetic irreversibility associated with spin disorder [32]. Interestingly, ZFC displays a broad peak in the temperature range of 80-240 K, which is significantly different from previous reports, where a sharp peak in ZFC has been observed at ~100 K [32-34]. Furthermore, the magnetization vanishes around 400 K, which corresponds to the Néel temperature of NCO [10]. Magnetic hysteresis (*M-H*) loop measurements has been recorded at various temperatures (5-300 K) under zero-field conditions (FIG. 3. (b)) reveal that the saturation magnetization ($M_s$) reaches a maximum value of approximately 0.86 $\mu_B$/f.u. at 5 K under an applied field of 5 T and decreases progressively with increasing temperature. The coercivity ($H_c$), extracted from the *M-H* loops (inset of FIG. 3. (b)), decreases systematically from ~2565 Oe at 5 K to ~16 Oe at 300 K. In previous reports, coercivity was found to be zero at a temperature where ZFC *M-T* shows a sharp peak, and this was correlated with the superparamagnetism [34]. The finite coercivity at 300 K and the broad peak in the ZFC *M-T* measurement rule out the superparamagnetism in our NCO at low temperature. Under zero-field-cooled conditions, the magnetic hysteresis loops remain symmetric across all temperatures between 5 K and 300 K. However, when the sample is cooled in the presence of an external magnetic field of -5 T (+5 T), the loops exhibit a noticeable shift toward the positive (negative) field direction, as shown in FIG. 3. (c). These opposite shifts under ±5 T cooling fields clearly indicate the presence of a finite EB in the system [12]. In addition to horizontal (magnetic field axis) shift, the hysteresis loops shift vertically (magnetization axis) as well. The shift along the magnetic field axis is typically due to unidirectional anisotropy associated with the rotatable pinned AFM spins at the boundary between two ferrimagnetic domains. On the other hand, the magnetization axis shift indicates the presence of uncompensated pinned spins at the same boundary, which do not rotate in response to the external field [35]. The EB field has been found to be 375 Oe at 5 K and decreases with increasing measurement temperature. The shift in the magnetization axis is 0.8 emu/g at 10 K and vanishes at room temperature. One can see that both the exchange bias field and magnetization axis shift become nearly zero as the temperature reaches 300 K (FIG. 3. (d)). The temperature at which the exchange bias effect vanishes is referred to as the blocking temperature [12], which is approximately 300 K for NCO. Although the ground state of single-phase NCO is ferrimagnetic, the observation of exchange bias under field-cooled conditions indicates the existence of antiferromagnetic or frustrated exchange coupling between neighbouring ferrimagnetic domains. The presence of such interfacial coupling is further supported by the observation of non-saturating M-H loops at various temperatures, even under magnetic fields up to 5 T. [36].

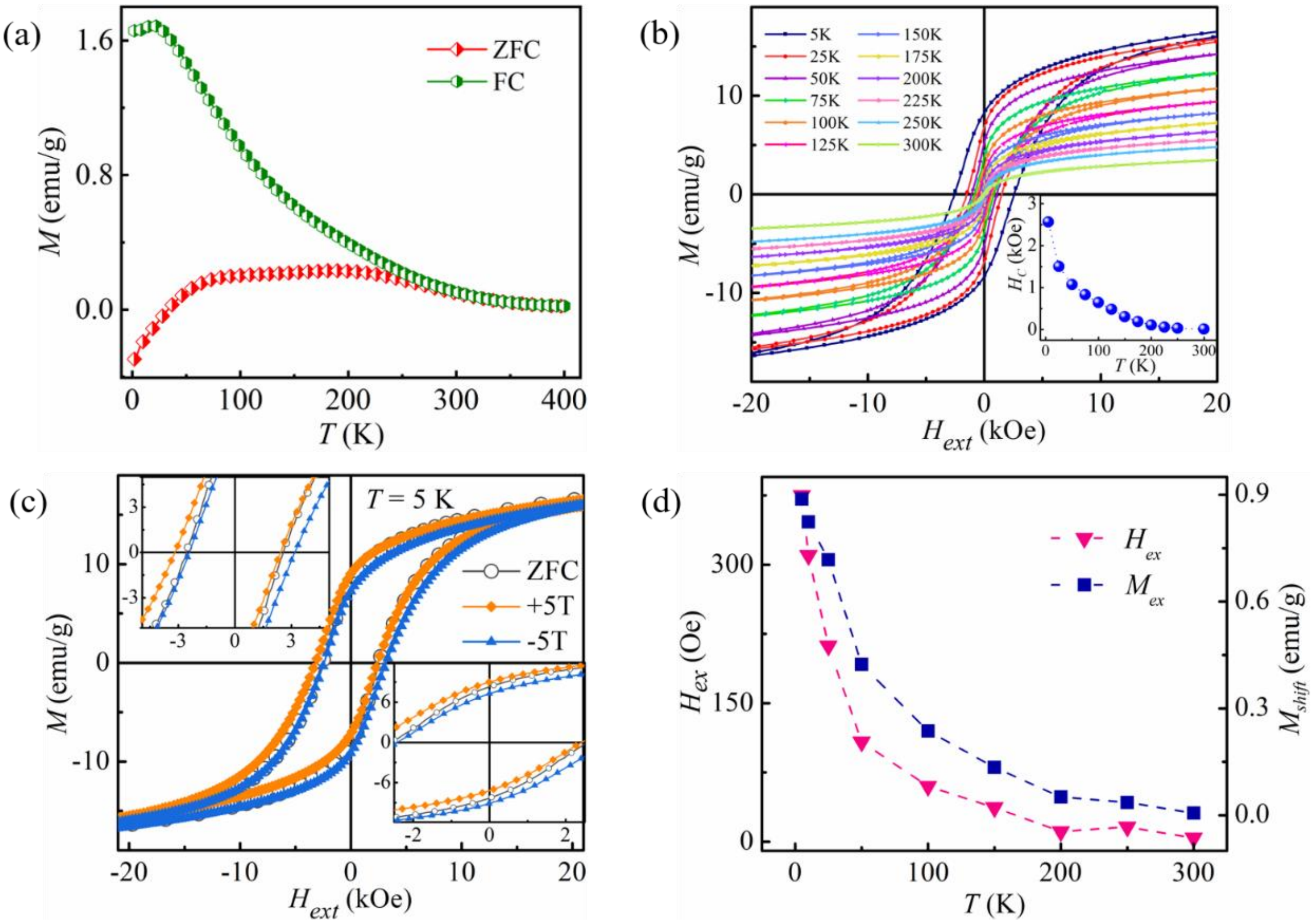

FIG 3. (a) ZFC and FC *M-T* curve of NCO measured under an applied field of 100 Oe. (b) *M-H* loops recorded at different temperatures, with the inset showing the variation of $H_c$ as a function of temperature. (c) *M-H* loops measured under ZFC and field-cooled (± 5 T) conditions, illustrating the horizontal and vertical loop shifts (shown in the upper-left and lower-right insets, respectively), indicative of exchange bias. (d) Temperature dependence of the exchange bias field ($H_{ex}$) and magnetization shift ($M_{shift}$), highlighting the disappearance of exchange bias near the blocking temperature (~300 K).

### 3.3. Transport properties

To gain further insight into the origin of the EB effect, the temperature dependence of the longitudinal resistivity ($\rho_{xx}$) and MR has been investigated. The $\rho_{xx}$ has been measured under zero magnetic field as a function of temperature from 5 to 300 K. As illustrated in FIG. 4. (a), the $\rho_{xx}$ exhibits a semiconducting behaviour, increasing sharply below 25 K. Upon the application of a high magnetic field of 9 T, the material retains its semiconducting nature; however, the $\rho_{xx}$ value has been significantly suppressed at low temperatures (below 100 K). This field-induced suppression of resistivity weakens progressively with increasing temperature and becomes negligible above 200 K. In disordered and polycrystalline semiconductors, charge transport is typically governed by various thermally activated hopping or tunnelling processes, depending on the different temperature regimes. The nearest-neighbour hopping (NNH) mechanism leads to a temperature dependence of resistivity given by [37-40],

$$\rho_{xx} = \rho_{NN}\ exp\left({E_a}/{k_B T}\right) \quad (1)$$

where $\rho_{NN}$ is the pre-factor and $E_a$ is the activation energy for hopping. In contrast, the variable range hopping mechanism exhibits the temperature dependence of [38, 40],

$$\rho_{xx} = \rho_{vh}\ exp\left({T_0}/{T}\right)^{1/4} \quad (2)$$

Here, $\rho_{vh}$ the hopping pre-factor and $T_0$ the Mott characteristic temperature are related to the density of localized states near the Fermi level. In case of intergrain tunnelling mechanisms (IGT) , the temperature dependence of $\rho_{xx}$ can be given as [41],

$$\rho_{xx} = \rho_{QT}\ exp\left({T_0'}/{T}\right)^{1/2} \quad (3)$$

Here, $\rho_{QT}$ is tunnelling pre-factor, and $T_0'$ denotes the characteristic temperature associated with the tunnelling barrier strength. Typically, $T^{-1/2}$ and $T^{-1/4}$ temperature dependence of $\ln(\rho_{xx})$ are observed at low (< 100 K) temperatures, whereas $T^{-1}$ dependency has been observed at high temperatures [38]. To elucidate the conduction mechanism responsible for the observed semiconducting behaviour of NCO, $\ln(\rho_{xx})$ is plotted as a function of $T^{-1}$, $T^{-1/4}$ and $T^{-1/2}$, as shown in FIG. 4. (b). In the low-temperature region (< 100 K), $\ln(\rho_{xx})$ exhibits a linear dependence on $T^{-1/2}$, indicating tunnelling-dominated transport, whereas in the intermediate temperature range (100-200 K), a linear relationship with $T^{-1/4}$ is observed, suggesting a variable range hopping (VRH) mechanism. In contrast, at higher temperatures (>200 K), $\ln(\rho_{xx})$ varies linearly with $T^{-1}$, consistent with a nearest-neighbour hopping (NNH) conduction process. The combined fitting using $T^{-1}$, $T^{-1/4}$ and $T^{-1/2}$ dependencies successfully reproduces the resistivity behaviour over the entire temperature range, as illustrated in the inset of FIG. 4. (a). Considering the pronounced suppression of resistivity at low temperatures, $\rho_{xx}$ has been further measured as a function of external magnetic field over the temperature range of 5-300 K.

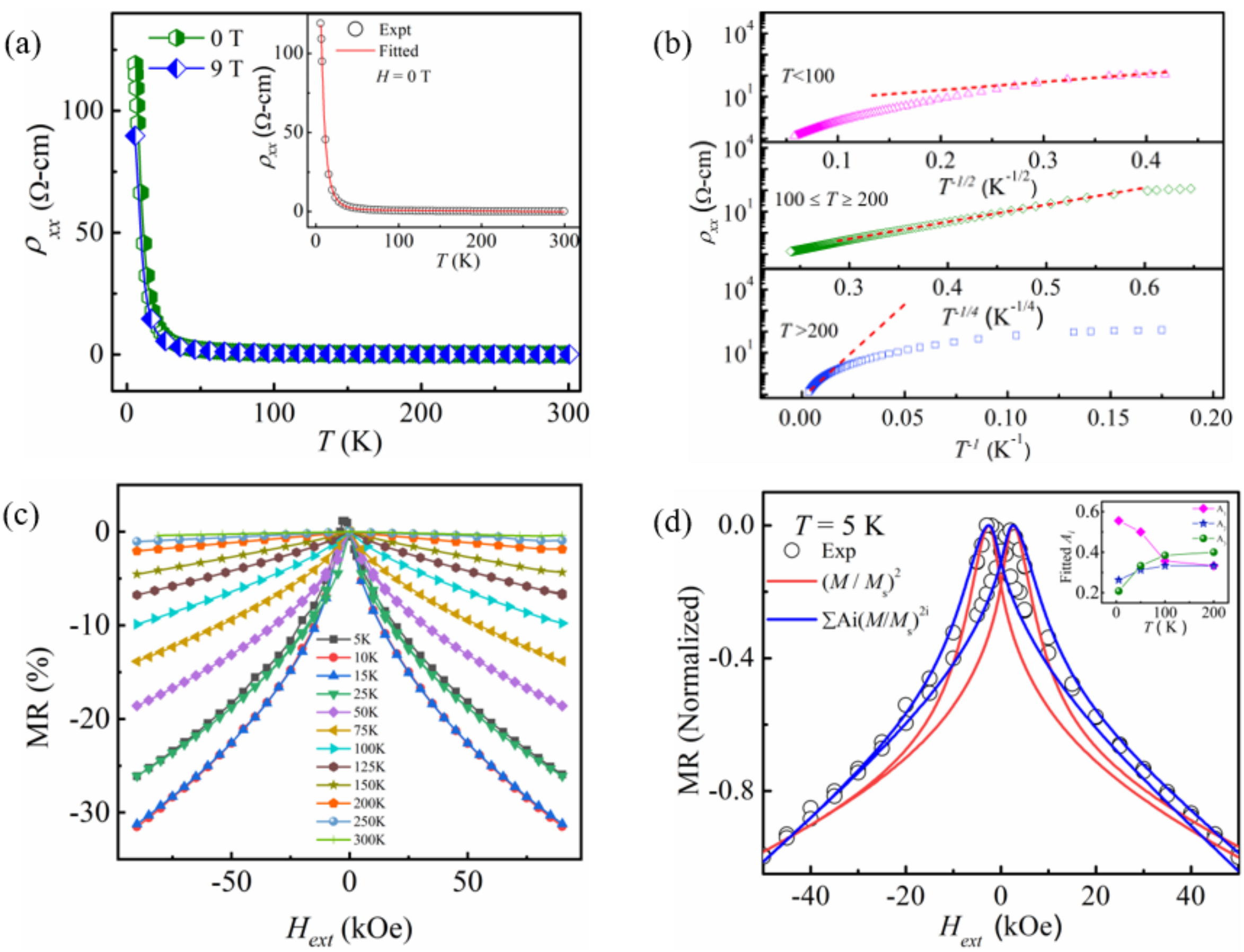

FIG. 4. (a) $\rho_{xx}$ *versus* $T$ plot at $H$ = 0 Oe and $H$ = 90 kOe, inset depicts the fitted data of $\rho_{xx}$ *versus* $T$ at $H$ = 0 $T$ using combined models of VRH, NNH, and IGT. (b) Fitting of $\rho_{xx}$ *versus* $T$ data in different temperature regions: T > 200 K (eqn. (1)), 200 K< T >100 K (eqn. (2)), and T< 100 K (eqn. (3)). (c) MR% of NCO measured between 5 K and 300 K. (d) Butterfly-shaped Low-field MR curve of NCO measured at 5 K, along with the corresponding fitting curves based on $(M/M_s)^2$ and $A_i(M/M_s)^{2i}$ models. The inset shows the temperature dependence of fitted parameters $A_i$ ($i$ =1,2,3).

The MR of NCO has been evaluated using the relation [42],

$$MR(\%) = \frac{[\rho_{xx}(H)-\rho_{xx}(0)]}{\rho_{xx}(0)} \times 100 \quad (4)$$

We observe that, with increasing temperature, the MR gradually decreases, as shown in FIG. 4. (c). The MR is negative and reaches its maximum value of ~-31.5% at 10 K, becoming negligible at 300 K. Notably, at low magnetic fields, the MR exhibits a butterfly-shaped hysteretic behaviour, with the peak position precisely matching the coercive field of the magnetic hysteresis loop as shown in FIG. 5.

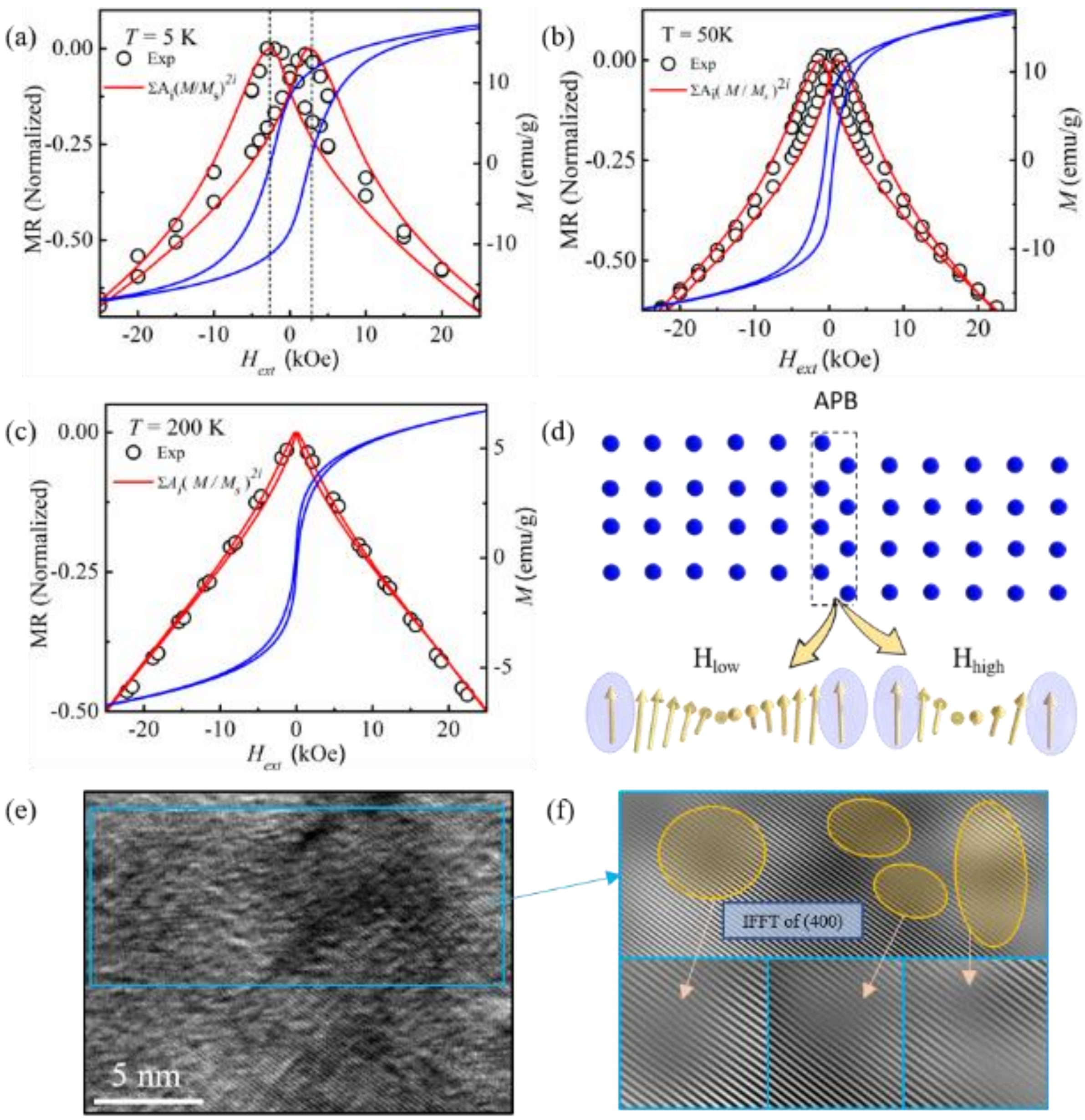

FIG. 5. (a-c) Hysteresis loops, measured MR data (open circles), and fitted curves $A_i(M/M_s)^{2i}$ obtained at 5 K, 50 K, and 200 K, respectively. (d) Schematic representation of antiphase boundary in the NCO system. (e) HRTEM image of NCO. (f) IFFT of the selected region of the HRTEM image representing APB.

The butterfly-shaped MR gradually weakens with increasing temperature and nearly vanishes around 200 K. Such butterfly-shaped MR behaviour has also been observed in spinel oxides, including $Fe_3O_4$ thin films [43]. In granular systems, the MR can be expressed as MR proportional to $\left(M/M_s\right)^{2i}$ where $M_s$ is the saturation magnetization and $i$ is an integer number [44]. Previously, it has been observed in nanocrystalline NCO that MR is proportional to $(M/M_s)^2$ at low magnetic field and low temperature [18]. In our study, we observe a strong deviation from the MR = $(M/M_s)^2$ behaviour, as displayed in FIG. 4. (d). Instead, we found that higher-order terms ($i$ = 2, 3) are crucial for reproducing the experimental observations of MR in the wide temperature range. The strength ($A_i$) of $(M/M_s)^2$, $(M/M_s)^4$, and $(M/M_s)^6$ are displayed in the inset of FIG. 4. (d). One can clearly see that with an increase in temperature, the $(M/M_s)^2$ contribution decreases and becomes of similar strength to the $(M/M_s)^4$ and $(M/M_s)^6$ terms. The butterfly shape, the peak of MR at coercivity of magnetic hysteresis, and the $(M/M_s)^{2i}$ dependence of MR clearly demonstrate the presence of antiphase boundary in the present NCO [44]. These boundaries occur in various systems, especially in spinel oxides, where the unit cell shifts by one-quarter, resulting in an interface with antiferromagnetic exchange coupling between two grains. Considering a grain equivalent to a ferromagnetic/ferrimagnetic domain, the antiferromagnetic exchange can be considered between two magnetic domains [45]. The behaviour of the antiphase boundary at low and high magnetic fields is schematically displayed in FIG. 5. (d). At a low magnetic field, spins across the APBs enhance spin disorder due to antiferromagnetic exchange coupling or magnetic frustration, which leads to a higher resistance. With increasing magnetic field, spins in these frustrated regions progressively align along the field direction, suppressing spin-disorder scattering and resulting in a pronounced negative MR. This effect is further strengthened at low temperatures due to reduced thermal fluctuations at the APBs. To confirm the presence of antiphase boundaries (APBs) in our sample, we further analyzed the high-resolution transmission electron micrograph. The Inverse Fast Fourier Transform of the [400] plane clearly displays the existence of APBs, which are mediated by the dislocation as shown in Fig. 5(e,f)[46].

## 4. Conclusion

A significant exchange bias is observed at 5 K (~375 Oe) when the NCO is cooled in the presence of an external magnetic field. We also observe both horizontal and vertical shifts in the magnetic hysteresis, which are related to unidirectional anisotropy and uncompensated pinned spins at the interface. Additionally, a large negative magnetoresistance has been observed at 10 K (~31.5%) and it shows hysteretic behaviour at low magnetic fields, with peak positions aligning with the coercive field of the magnetic hysteresis. The scaling of magnetoresistance with $(M/M_s)^{2i}$ indicates the presence of an antiphase boundary in the NCO nanoparticle sample. High-resolution transmission micrograph confirms the formation of APB. This antiphase boundary leads to an interface with antiferromagnetic exchange coupling between two ferrimagnetic domains, resulting in a significant exchange bias. This study provides a path for utilizing the interface of a single material for exchange bias, which can be widely used in various spintronic applications, such as magnetic random-access memory where magnetic-field-free switching can be observed using this interfacial exchange bias.

**Funding**

This work has been funded and supported by the IIT Guwahati Start-up Grant and the DST INSPIRE Faculty Fellowship under project no. (DST/INSPIRE/04/2022/003287), Government of India.

**CRediT authorship contribution statement**

**Biswanath Pramanik**: Writing – original draft, Methodology, Investigation, Formal analysis, Data Curation**. Pushpesh Pathak**: Writing – original draft, Investigation. **Binoy K. Hazra**: Writing–review & editing, Supervision, Resources, Project administration, Funding acquisition, Conceptualization.

**Declaration of Competing Interest**

The authors declare that they have no known competing financial interests or personal relationships that could have appeared to influence the work reported in this paper.

**Data Availability**

Data will be made available on reasonable request.

**Acknowledgements**

This research was supported by the IIT Guwahati Start-up Grant and the DST INSPIRE Faculty Fellowship (DST/INSPIRE/04/2022/003287), Government of India. B. Pramanik gratefully acknowledges the Ministry of Education, India, for the fellowship. The authors also acknowledge the Central Instrument Facility and the Department of Physics, IIT Guwahati, for access to instrumentation procured through the DST project (SR/FST/PSII-037/2016).